\documentclass{emulateapj}

\usepackage{amsmath}

\newcommand{\about}{$\sim\!\!$~}

\newcommand{\be}{\begin{displaymath}}
\newcommand{\ee}{\end{displaymath}}

\def\lsim{\hbox{\rlap{\raise 0.425ex\hbox{$<$}}\lower 0.65ex\hbox{$\sim$}}}
\def\gsim{\hbox{\rlap{\raise 0.425ex\hbox{$>$}}\lower 0.65ex\hbox{$\sim$}}}
\def\arcmin{\hbox{$^\prime$}}

\newcommand{\msun}{M$_\sun$}

\newcommand{\halpha}{H$\alpha$}

\newcommand{\hbeta}{H$\beta$}
\newcommand{\hgamma}{H$\gamma$}

\newcommand{\kms}{km~s$^{-1}$}
\newcommand{\perMpc}{Mpc$^{-1}$}
\newcommand{\ergps}{ergs~s$^{-1}$}

\newcommand{\ig}{SN~2001ig}

\shorttitle{Optical Spectroscopy of SN~2001ig}
\shortauthors{Silverman et al.}

\begin{document}

\title{Optical Spectroscopy of the Somewhat Peculiar Type~IIb Supernova 2001ig}

\author{Jeffrey M. Silverman\altaffilmark{1},
  Paolo Mazzali\altaffilmark{2,3},
  Ryan Chornock\altaffilmark{1},
  Alexei V. Filippenko\altaffilmark{1},
  Alejandro Clocchiatti\altaffilmark{4},
  Mark M. Phillips\altaffilmark{5}, 
  Mohan Ganeshalingam\altaffilmark{1}, and
  Ryan J. Foley\altaffilmark{1,6,7}
}

\altaffiltext{1}{Department of Astronomy, University of California,
Berkeley, CA 94720-3411; JSilverman@astro.berkeley.edu.}

\altaffiltext{2}{Max-Planck Institut f\"{u}r Astrophysik,
  Karl-Schwarzschild-Strasse 1, 85748 Garching, Germany.}

\altaffiltext{3}{Istituto Naz. di Astrofisica-Oss. Astron., vicolo
dell'Osservatorio, 5, 35122 Padova, Italy.}

\altaffiltext{4}{Pontificia Universidad Cat\'olica de Chile, 
Departamento de Astronom\'ia y Astrof\'isica, Casilla 306, Santiago 22, Chile.}

\altaffiltext{5}{Las Campanas Observatory, Casilla 601, La Serena, Chile.}

\altaffiltext{6}{Harvard-Smithsonian Center for Astrophysics, Cambridge,
MA 02138.}

\altaffiltext{7}{Clay Fellow.}

\begin{abstract} 
Here we present previously unpublished optical spectra of supernova
(SN) 2001ig, a Type~IIb SN, from about a week
after explosion until nearly one year later. The earliest
spectrum consists of only a few broad absorption features,
but soon more common Type~II SN features including
hydrogen P-Cygni profiles and helium absorption become apparent.
At later times,
as the H features fade and the \ion{He}{1} absorption becomes
more prominent, we observe the SN to transition from a Type~II to a Type~Ib.
Finally, observations after 250 days past explosion show a
nebular-phase SN spectrum with one of the largest magnesium
to oxygen intensity ratios ever seen.
Additionally, we present 
models of the late-time spectra which indicate that the 
inner ejecta consist of \about1.15~\msun\ of material, most of which (by mass) is
in the form of oxygen, with \about0.13~\msun\ of $^{56}$Ni and
essentially no hydrogen.
\end{abstract}



\section{Introduction}\label{s:intro}

It is thought that most high-mass stars ($\gtrsim8$~\msun) end their lives
as core-collapse supernovae (SNe) \citep[e.g.,][and references therein]{Woosley05}.
These SNe are divided spectroscopically into subgroups based on the
strength of H and He in their optical spectra. The progression from Type~II to Ib to Ic
corresponds to SNe having both H and He, to just He, to neither H nor He
present in their spectra
\citep[for a review, see][]{Filippenko97}. The class of
Type~IIb SNe (SNe~IIb) have spectra which undergo a transition from Type~II to Type~Ib as
their H features fade with time. It is likely that SNe~IIb, Ib, and Ic
undergo significant mass loss,
due either to stellar winds and eruptions or to 
mass transfer to a binary companion, before explosion.


\ig\ was discovered in the nearby late-type spiral galaxy \object{NGC 7424}
by \citet{Evans01} on 2001~Dec.~10.43 (UT dates are used throughout
this paper). It is located at $\alpha_{\rm J2000} = 
22^{\rm h}57^{\rm m}30\fs 74$ and 
$\delta_{\rm J2000} = -41\degr02\arcmin26\farcs1$ \citep{Ryder01}.
An explosion date of 2001~Dec.~3 has 
been estimated by \citet{Ryder04} from modeling the radio light curve, and we
use this as the reference date for our spectral observations
throughout this paper. A spectrum was obtained within two days of
detection of the SN and showed
only a few broad absorption features. However, the spectrum from Las 
Campanas Observatory on 2001~Dec.~13 \citep{Phillips01} revealed
similarities to the Type~IIb SNe~1987K \citep{Filippenko88} and
1993J \citep{Filippenko93}.


Over the following month, spectra from the European Southern Observatory
\citep{Clocchiatti01,Clocchiatti02} showed \ig\ to evolve in a manner
similar to both SNe~IIb~1987K and 1993J,
as the hydrogen lines became
weaker and showed the complex shape characteristic of the transitional stage
\citep{Filippenko94}. By Oct.~2002, the predicted transition to a
Type~Ib SN in the nebular phase was complete \citep{Filippenko02}.
At the time it was also noted that the \ion{Mg}{1}] $\lambda$4571 feature
was the strongest (relative to [\ion{O}{1}] $\lambda\lambda$6300, 6364)
ever seen in a SN. Our data complement the optical spectropolarimetry and total-flux
spectra that were presented and analyzed by \citet{Maund07}.

In addition to optical data, X-ray images were
obtained with the Advanced CCD Imaging Spectrometer-S (ACIS-S) instrument
on the {\it Chandra} X-ray Observatory on 2002~May~22, showing a
0.2--10.0~keV luminosity of \about$10^{38}$~\ergps\ \citep{Schlegel02}. \ig\
was also detected
at radio frequencies; for further information on the radio data and
analysis, see \citet{Ryder04}.

Here we present and analyze optical spectroscopic data
as well as two spectral models for \ig\ at late times.  In \S\ref{s:obs}
we describe our observations and data
reduction, and in \S\ref{s:analysis} we discuss our analysis of the
spectra. Our model spectra and their implications are presented in
\S\ref{s:model}. We summarize our conclusions in \S\ref{s:conclusions}.


\section{Observations and Data Reduction}\label{s:obs}

Beginning one week after explosion, optical spectra were obtained using
the LDSS-2 spectrograph \citep{Mulchaey01} mounted on the
Magellan Baade 6.5-m telescope,
the red arm of the dual imager/spectrograph EMMI
\citep{Dekker86} mounted on the European Southern Observatory (ESO) 3.6-m New Technology
Telescope, the Low-Resolution Imaging Spectrometer \citep[LRIS;][]{Oke95} mounted on the Keck I 10-m
telescope, and the dual imager/spectrograph EFOSC2 \citep{Buzzoni84} mounted on the
ESO 3.6-m telescope. Our last spectral observation occurred
on 2002~Nov.~8, approximately 340 days after explosion.
All observations used a $0.7''$--$1.0''$ wide slit and were aligned along the parallactic angle
to reduce differential light losses \citep{Filippenko82}.
Table~\ref{t:obs}
summarizes the spectral data of \ig\ presented in this paper.

\begin{deluxetable*}{lrllrcr}
\tablecaption{Journal of Observations\label{t:obs}}
\tablewidth{0pt}
\tablehead{
\colhead{UT Date} &
\colhead{Day\tablenotemark{a}} &
\colhead{Telescope\tablenotemark{b}} &
\colhead{Range} &
\colhead{Res.} &
\colhead{Airmass\tablenotemark{d}} &
\colhead{Exp.} \\
 & & &
\colhead{(\AA)} &
\colhead{(\AA)\tablenotemark{c}} & &
\colhead{(s)} }
\startdata
2001 Dec.\ 11  & 8  &   LDSS2 &   3800--9000  & 13  &   1.07 &  $2\times90$ \\
2001 Dec.\ 13   &   10  &  LDSS2  &  5000--10000 & 13  &   1.18  & $2\times60$ \\
2001 Dec.\ 16   &   13  &  EMMI   &  3980--9400  &  9 &  1.64  & $3\times60$ \\
2001 Dec.\ 23   &   20  &  LRIS    & 3400--10300  & 7  &  2.49  & 60 \\
2002 Jan.\ 13   &   41  &  EFOSC2  & 3360--10300  & 9 &  1.87 &  $3\times90$ \\
2002 Jan.\ 17   &   45  &  LRIS   &  3520--10000  & 7  &  4.09 &  360 \\
2002 Jan.\ 18    &  46  &  LRIS   &  3500--5380   & 3 &   4.99  & 450 \\
2002 Jan.\ 18   &   46  &  LRIS   &  5710--7000  &  2 &   4.99  & 450 \\
2002 Jan.\ 29    &  57  &  LDSS2  &  3600--10000 & 13 &  2.61  & 90 \\
2002 Aug.\ 16\tablenotemark{e}     &    256  & FORS1   & 4000--8600 &  12 &  1.04  & $4\times1200$ \\
2002 Oct.\ \phn8   &   309  & LRIS    & 3150--9420  &  6  &  2.18  & $2\times600$ \\
2002 Nov.\ \phn8   &   340  & LRIS    & 3100--9410   & 6   & 2.05  & 900 \\
\enddata
\tablenotetext{a}{Days since the explosion date of 2001~Dec.~3
  \citep[estimated by][from the radio light curve]{Ryder04}.}
\tablenotetext{b}{{L}DSS2 = Magellan Baade 6.5-m/LDSS-2 spectrograph \citep{Mulchaey01};
EMMI = European Southern Observatory (ESO) 3.6-m New Technology Telescope/red arm of the dual
imager/spectrograph EMMI \citep{Dekker86}; LRIS = Keck I
10-m/Low-Resolution Imaging Spectrometer \citep{Oke95};
EFOSC2 = ESO 3.6-m/dual imager/spectrograph EFOSC2 \citep{Buzzoni84};
FORS1 = ESO 8.2-m Very Large Telescope/FORS1 spectrograph \citep{Appenzeller98}.}
\tablenotetext{c}{Approximate spectral resolution, \AA.}
\tablenotetext{d}{Airmass at midpoint of exposures.}
\tablenotetext{e}{Spectrum previously published by \citet{Maund07}.}
\end{deluxetable*}

All spectra were reduced using standard techniques
\citep[e.g.,][]{Foley03}.  Standard CCD processing and spectrum
extraction for the LRIS data were completed with
IRAF\footnote{IRAF: the Image Reduction and Analysis Facility is
distributed by the National Optical Astronomy Observatory, which is
operated by the Association of Universities for Research in Astronomy
(AURA) under cooperative agreement with the National Science
Foundation.}, and the data were extracted with the optimal
algorithm of \citet{Horne86}.  
We obtained the wavelength scale from low-order polynomial fits to
calibration-lamp spectra.
Small wavelength shifts were then applied to the
LRIS data after cross-correlating a template sky to the night-sky lines that
were extracted with the SN.
Using our own IDL routines, we fit spectrophotometric
standard-star spectra to the LRIS data in order to flux calibrate our spectra
and to remove telluric lines \citep{Wade88, Matheson00}.

Each EMMI observation was reduced independently and the final spectra were
combined with a pixel-rejection algorithm based on the mean.


\section{Spectral Analysis}\label{s:analysis}
\subsection{Early-Time Spectra}\label{ss:early}

\begin{figure*}
\epsscale{0.7}
\rotatebox{90}{
\plotone{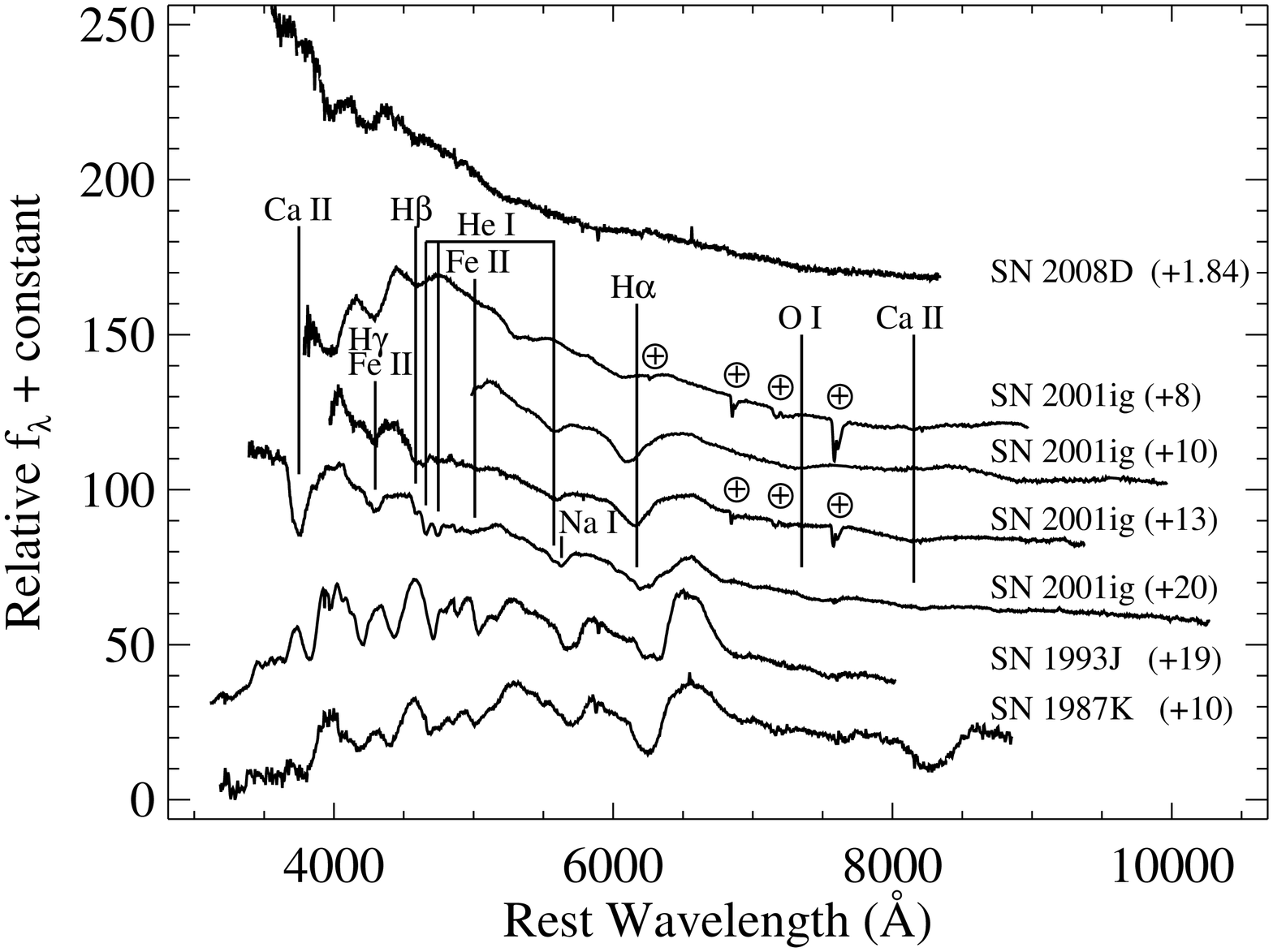}}
\caption{Spectra of \ig\ (and three comparison SNe)
with some line identifications.  The top spectrum is SN~2008D
\citep{Modjaz09} and has been dereddened by
$E\left(B-V\right)_{\rm Host}=0.6$~mag. The next four spectra, of \ig, 
from top to bottom were obtained on 2001~Dec.~11, 13, 16, and
23; ages shown are relative to the adopted explosion date of
2001~Dec.~3. The bottom two spectra are
SN~1993J \citep{Filippenko94} and SN~1987K
\citep{Filippenko88}. Recession velocities of 2100~\kms\
\citep{Modjaz09}, 939~\kms\ \citep{Koribalski04},
$-34$~\kms, and 805~\kms\ \citep[both from][]{DeVaucouleurs91} have been
removed from the spectra of SNe~2008D, 2001ig,
1993J, and 1987K, respectively. Telluric features are indicated by 
the $\earth$ symbol.}\label{f:early_spec}
\end{figure*}

We present our early-time spectra of \ig\ in Figure~\ref{f:early_spec}; all 
have had the recession velocity of \object{NGC 7424} removed
\citep[939~\kms, from][]{Koribalski04}. 
The first spectrum of \ig, taken on 2001~Dec.~11 (8 days past explosion),
shows only a few strong, broad
absorption features.
We conjecture that the feature near 4000~\AA\ is \hgamma\ and the
feature near 4300~\AA\ is
a blend of \ion{Fe}{2}
lines (near a rest wavelength of 4570~\AA)
blueshifted by \about22,500~\kms\ and \about18,200~\kms, respectively.
Unfortunately, this identification is somewhat dubious since the \hgamma\
feature appears to be unusually strong.

Instead, we note that this pair of features resembles the ``W'' feature
seen in very early-time spectra of the Type~II SN~2005ap \citep{Quimby07}
and the Type~Ib SN~2008D \citep[][a spectrum of which is reproduced in
Figure~\ref{f:early_spec}]{Malesani09,Modjaz09}.  The features in both
objects were attributed to a combination of \ion{C}{3}, \ion{N}{3},
and \ion{O}{3} at blueshifted velocities of \about21,000~\kms\ and \about30,000~\kms,
respectively.  If we assume that these are the correct line
identifications, then they are blueshifted by \about26,000~\kms\ in
the day 8 spectrum of \ig.  However, there is some
uncertainty in the bluer of these two features since its left wing is
distorted somewhat by the noise at the bluest end of our spectrum.
It should also be noted that \citet{Mazzali08} found no evidence for
the ``W'' feature in their earliest spectrum of SN~2008D.

The third feature (near 4600~\AA) can be decomposed into two Gaussian profiles
centered at \about4498~\AA\ and \about4596~\AA, which could
be a blend of \hbeta\ and \ion{He}{1} $\lambda\lambda$4921, 5015
blueshifted by \about23,200~\kms\ and \about23,300~\kms, respectively. 
At this epoch
\ig\ defies simple classification \citep[as pointed out by][]{Phillips01}.

There is also weak, but broad,
absorption at \about5300~\AA\ and 5420~\AA, the latter of which could be
\ion{He}{1} $\lambda$5876 
blueshifted by \about24,100~\kms. 
In addition, the absorption
near 6060~\AA\ is most likely \halpha\ blueshifted by \about24,000~\kms.

We measure a blackbody-fit temperature of
\about14,000~K for the continuum of this observation.
In comparison, the derived early-time blackbody temperatures of
SNe~1993J and 2008D were \about12,000~K \citep{Clocchiatti95} and
13,000~K \citep{Modjaz09}, respectively, which match nicely with \ig.

In our highest-resolution spectrum, taken on 2002~Jan.~18 (46 days
after explosion; see \S\ref{ss:transition} for more on this
observation), we measure an upper limit to the equivalent width (EW)
of the \ion{Na}{1}~D line in \object{NGC 7424} of \about0.1~\AA.  This
corresponds to a host-galaxy reddending of $E\left(B-V\right) \lesssim
0.02$~mag \citep{Munari97}, which is comparable to the Galactic
reddening of $E\left(B-V\right)=0.011$~mag from
\citet{Schlegel98}. The low host reddening for \ig\ is reasonable since the
SN occurred on the outskirts of the galaxy. Since
these numbers are both quite small, we ignore reddening when plotting
our spectra.   We also note that \citet{Maund07}
required a total reddening somewhat in
excess of the Galactic value plus the value we calculate for the host
to explain the polarization they observe.

In all other early-time spectra, we detect
\ion{He}{1} $\lambda$5876 and P-Cygni profiles of \halpha\ (with
the absorption component greatly dominating over the emission component).
The observation from 2001~Dec.~13, for example, shows the \ion{He}{1} feature blueshifted
by \about16,000~\kms.
In addition, \citet{Phillips01} mention that there are broad absorption
features at 4080~\AA\ and 4580~\AA\ in the observation from
2001~Dec.~13,
and it is possible that they are \hgamma\ and
\hbeta, respectively.
However, the observation lacked acceptable blue-side flatfields
and proper wavelength-calibration files, so data at wavelengths
$\lesssim$5000~\AA\ cannot be completely trusted.
We also identify
the very weak, broad absorption near 7300~\AA\ as \ion{O}{1} $\lambda$7774.

By 13 days after the explosion 
many more spectral features have appeared.
Still present are the P-Cygni profile of \halpha\  
and the \ion{He}{1} $\lambda$5876 line. 
\citet{Maund07} point out that \halpha\ at this time is likely to be blended with
\ion{He}{1} $\lambda$6678.
No longer visible in this spectrum is the aforementioned ``W'' feature
that probably came from a combination of \ion{C}{3}, \ion{N}{3},
and \ion{O}{3}.  This is again similar to SN~2008D, whose ``W''
disappeared in the span of \about1 day \citep{Modjaz09}.
However, there {\it are} still features at about 4130~\AA\ and
4287~\AA, and we associate these with possible absorption due to \hgamma\ 
and \ion{Fe}{2}, respectively.

The feature near 4600~\AA\ is most likely \hbeta\  
while the noisy dips just below 4800~\AA\ are probably due to
a combination of \ion{He}{1}
$\lambda\lambda$4921, 5015 and \ion{Fe}{2} $\lambda\lambda$4924, 5018.
Also seen is weak absorption likely due to \ion{Fe}{2} $\lambda$5169 near
5030~\AA\ 
and quite broad, weak absorption from the \ion{Ca}{2} near-infrared (IR) 
triplet around
8150~\AA.

On 2001~Dec.~23, we acquired the fifth spectrum from the top of Figure~\ref{f:early_spec}.  
Again, we detect the \halpha\ P-Cygni line, although now it has a far
more complex profile with a larger full width at half-maximum
intensity (FWHM) than previously seen.  This implies that
the \halpha\ feature is now likely becoming blended with \ion{Si}{2} $\lambda$6355. 
The feature near 5600~\AA\ in this spectrum clearly shows two local
minima which are probably
due to \ion{He}{1} $\lambda$5876 and \ion{Na}{1}~D. 
Also in this observation we detect distinct \hbeta, \ion{He}{1} $\lambda$4921, 
and \ion{He}{1} $\lambda$5015 absorption, in addition to
absorption from the \ion{Ca}{2} near-IR triplet around 8200~\AA. 
Finally, extremely strong absorption due to \ion{Ca}{2}~H\&K can
now be seen near 3750~\AA.

It has been pointed out that in many ways the first few spectra of \ig\ resemble
those of SNe~1987K and 1993J at early times
\citep{Phillips01,Clocchiatti01}.  Examples of both at similar ages are plotted
for comparison in Figure~\ref{f:early_spec} and there do appear to be
quite a few similarities among these objects.  However, it should be
noted that \halpha\ (both the absorption and
emission components) is weaker in \ig.  In addition, the overall
continuum shape of \ig\ does not change much during this
two-week period. 

Figure~\ref{f:early_spec} also indicates that \ig\
has a bluer continuum than either SN~1987K or SN~1993J; however, this could
be due to reddening.  Both SNe~1987K and 1993J are
plotted with no reddening correction, which is approximately consistent with
observations \citep{Richmond94}, but there is also evidence that the color excess of
SN~1993J is as high as 0.25--0.32~mag \citep{Richmond94,Clocchiatti95}.

\subsection{Transition Spectra}\label{ss:transition}

\begin{figure*}
\epsscale{0.7}
\rotatebox{90}{
\plotone{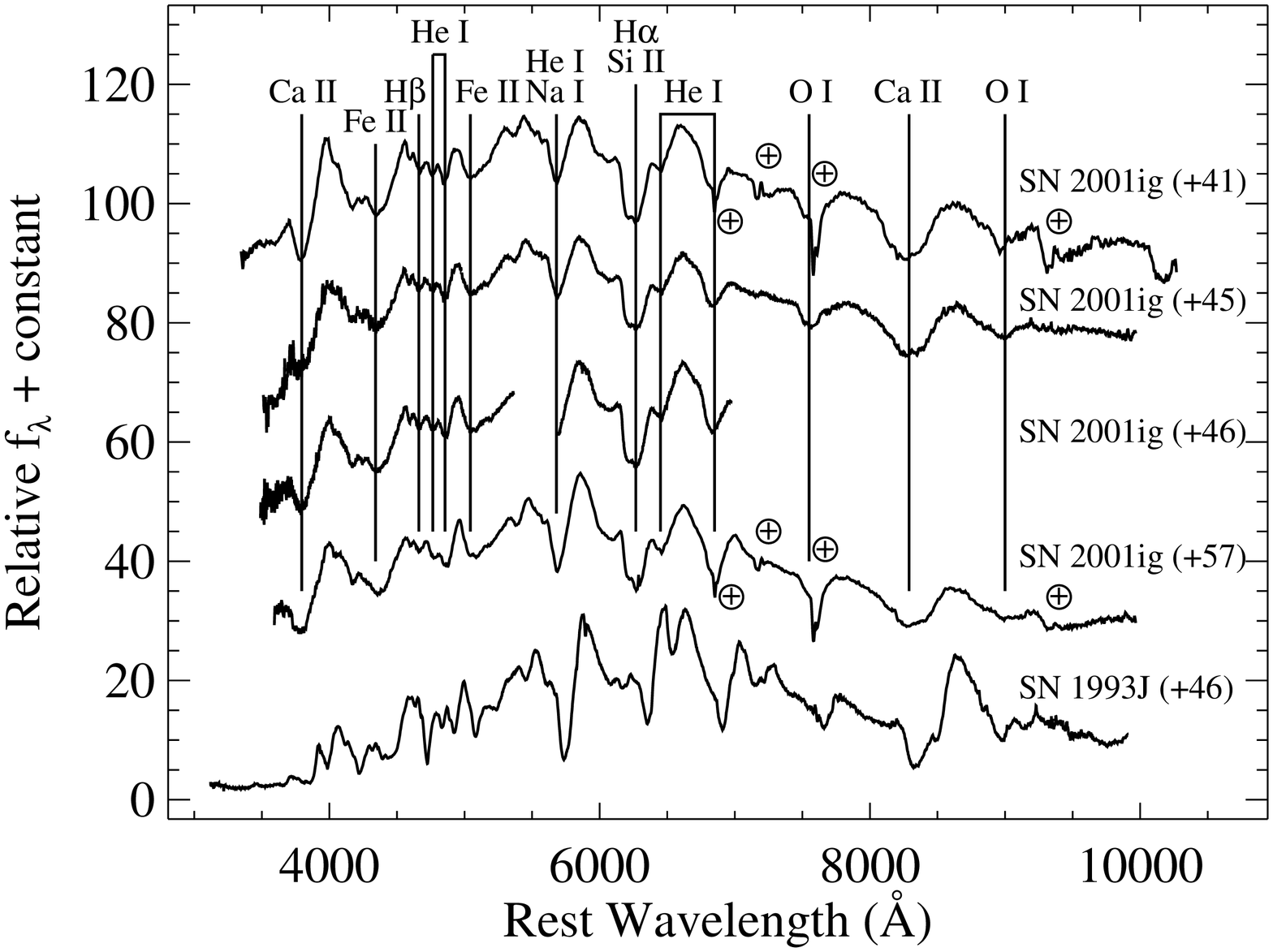}}
\caption{Spectra of \ig\ (and a comparison SN IIb) with some line identifications.  From top to bottom,
the spectra were obtained on 2002~Jan.~13, 17, 18, and 29; ages shown 
are relative to the adopted explosion date of 2001~Dec.~3.
The bottom
spectrum is SN~1993J
\citep{Filippenko94}. Recession velocities have been removed from the spectra.
Telluric features are indicated by the $\earth$ symbol.}\label{f:mid_spec}
\end{figure*}

Our spectra from \about5.5 weeks past explosion to \about8 weeks past explosion are
presented in Figure~\ref{f:mid_spec}.  The most obvious change in the
spectra of \ig\ from
Figure~\ref{f:early_spec} to Figure~\ref{f:mid_spec} is that the flux
in the blue part of the continuum 
has decreased significantly; \ig\ now looks very much like SN~1993J in
its overall continuum shape. In addition, SNe~2001ig and 1993J look
closely similar in the range \about4000--6000~\AA, much of which
is a forest of \ion{Fe}{2} blends (with some hydrogen Balmer and
\ion{He}{1} lines as well).
As noted by \citet{Clocchiatti02}, the main difference at this time
between SNe~2001ig and 1993J is that \ig\ has stronger
\ion{Ca}{2}~H\&K lines, weaker \ion{H}{1} and \ion{He}{1} lines,
and faster expansion velocities.


Also mentioned by \citet{Clocchiatti02}, by 2002~Jan.~13 (the top spectrum in
Figure~\ref{f:mid_spec}), \ig\ has begun its transition from Type~II to Type~Ib.
This transformation is marked by the decrease in strength of \halpha\
and \hbeta\ while the \ion{He}{1} spectral features increase in strength.  The most notable
cases of this are the dramatic
appearances of the \ion{He}{1} $\lambda$6678 absorption feature, which appears as a notch
on the red wing of the \halpha\ absorption profile, and strong
\ion{He}{1} $\lambda$7065
absorption, which unfortunately coincides with a telluric absorption band
in the day 41 and 57 spectra.
This transformation continues during the subsequent two weeks of observations
(i.e., the rest of the \ig\ spectra in Figure~\ref{f:mid_spec}).

The minimum of the \halpha\ absorption
feature is becoming more complex and appears to have
developed a doubled-troughed profile. This is to be expected since at
this time \halpha\ is being heavily
blended with \ion{Si}{2} $\lambda$6355.
In addition, as was pointed out by \citet{Maund07}, the \ion{He}{1}
$\lambda$5876 feature has become more prominent and the red wing of
its absorption now shows a notch due to \ion{Na}{1} D absorption.

Also, by this time the \ion{Ca}{2}~H\&K feature has developed a P-Cygni profile with
quite a strong emission component.  In addition, the \ion{Ca}{2}
near-IR triplet absorption
has strengthened dramatically.  The previously identified absorption
from \ion{Fe}{2} near 4300~\AA\ and 5000~\AA\ 
becomes stronger, broader, and more complex with time.
This is likely due to line blanketing
which was also seen in SN~2002ap \citep{Foley03}.

The broad absorption near 9000~\AA\ could be \ion{O}{1} $\lambda$9266
at about $-9000$~\kms.
In addition, strong, relatively broad absorption
from \ion{O}{1} 
$\lambda$7774 is again seen in these observations, but at
roughly $-8000$~\kms.
However,
this feature coincides with a telluric absorption band
in the day 41 and 57 spectra,
making the velocity determination difficult and uncertain.

\subsection{Late-Time Spectra}\label{ss:late}

\begin{figure*}
\epsscale{0.7}
\rotatebox{90}{
\plotone{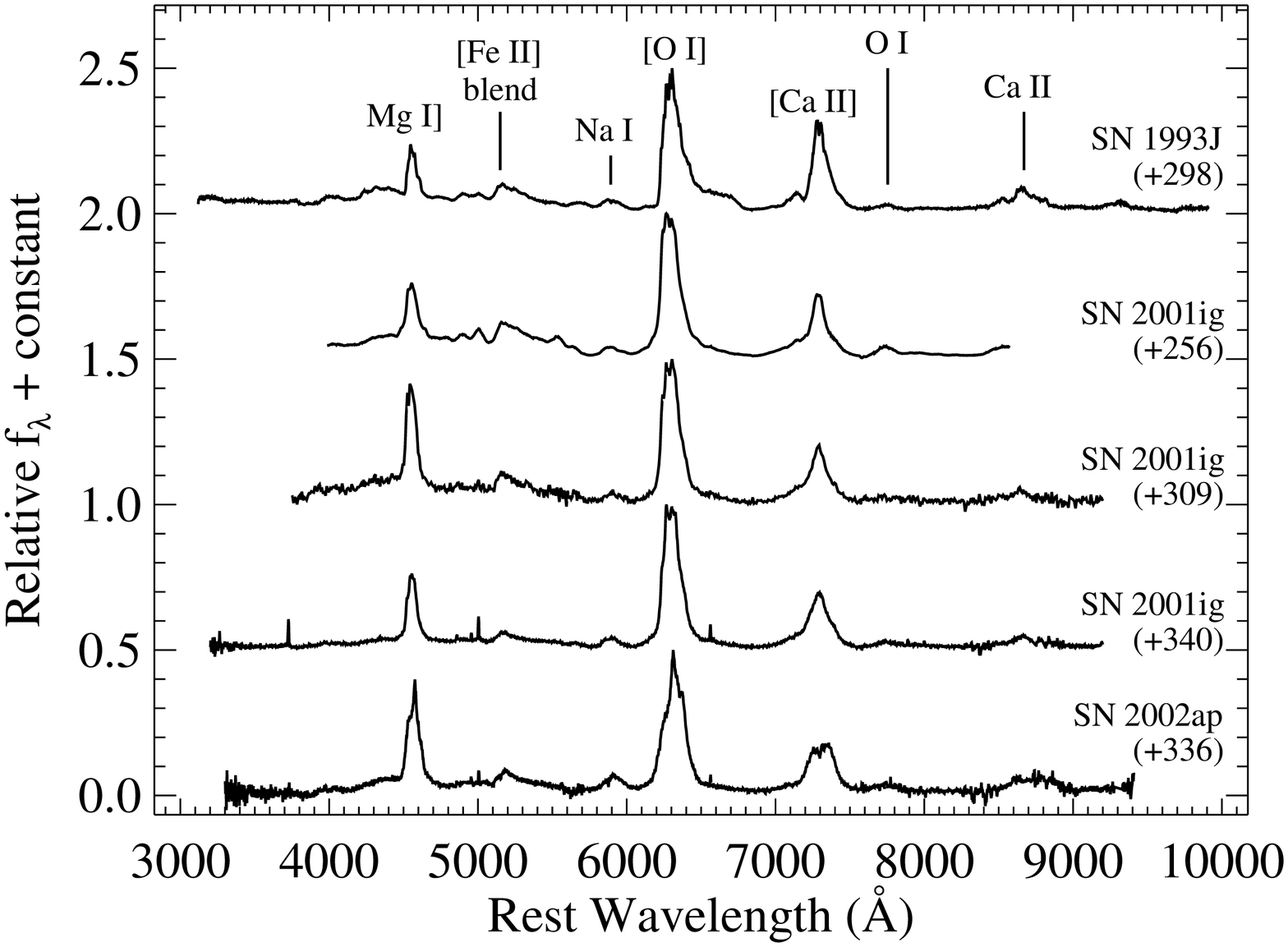}}
\caption{Spectra of \ig\ (and two comparison
SNe) with some line identifications.  The top spectrum is SN~1993J
\citep{Filippenko94}. The next three spectra
are \ig\ and were obtained on 2002~Aug.~16 \citep[from][Figure 2,
reproduced with permission]{Maund07}, Oct.~8,
and Nov.~8; ages shown are
relative to the adopted explosion date of 2001~Dec.~3. 
The bottom spectrum is the Type~Ic SN~2002ap \citep{Foley03}.
Recession velocities have been removed from the spectra
\citep[657~\kms\ for SN~2002ap, from][]{Lu93}.}\label{f:late_spec}
\end{figure*}

A late-time spectrum from \citet{Maund07} (256 days past explosion) and our
late-time spectra (309 and 340 days past explosion) are
presented in Figure~\ref{f:late_spec} and are compared with similar
late-time spectra of SNe~1993J and 2002ap.  \ig\ has reached the
nebular phase by this time and has quite strong [\ion{O}{1}]
$\lambda\lambda$6300, 6364 and \ion{Mg}{1}] $\lambda$4571 emission. These and
other common nebular SN features are marked in Figure~\ref{f:late_spec};
the features of SNe~2001ig, 2002ap, and 1993J
are all similar, but the relative
strengths of some of the emission lines differ. 

One of the most notable differences among the objects 
is that SN~1993J exhibits obvious \halpha\
emission by day 298, appearing as a shoulder on the red wing of [\ion{O}{1}]
$\lambda\lambda$6300, 6364. \citet{Filippenko94} and
\citet{Matheson00} showed that the \halpha\ feature gradually 
strengthened from this time onward in SN~1993J, whereas in
spectra at similar epochs of \ig\ shown in Figure~\ref{f:late_spec}
there is no obvious \halpha\ emission.  For SN~1993J the hydrogen at
late times was attributed to interaction with circumstellar gas, but
in \ig\ we see no evidence for such an interaction.

\begin{figure*}
\epsscale{1.1}
\rotatebox{90}{
\plotone{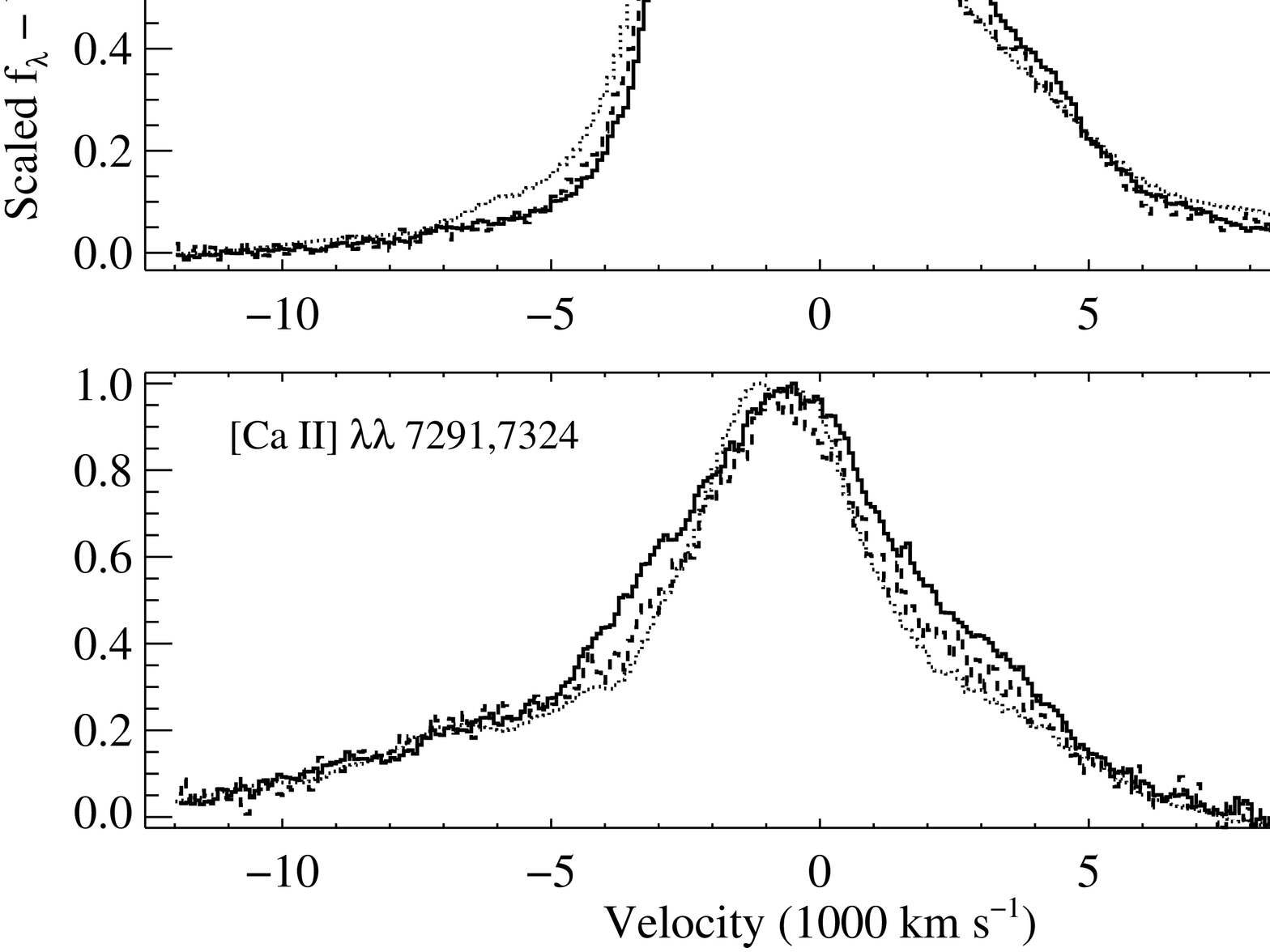}}
\caption{The profiles of \ion{Mg}{1}] $\lambda$4571 (top),
[\ion{O}{1}] $\lambda\lambda$6300, 6364 (middle), and
[\ion{Ca}{2}] $\lambda\lambda$7291, 7324 (bottom)
in \ig\ at three epochs, relative to the systemic velocity of the host
galaxy NGC 7424.
Ages relative to the adopted explosion date of 2001~Dec.~3 are shown.  To create
these figures, a local linear background was subtracted from each feature
and the maxima of the features were all scaled to 1.0.}\label{f:OMG_comparison}
\end{figure*}

Figure~\ref{f:late_spec} also shows that the blended [\ion{Fe}{2}]
emission near 5200~\AA\ is weak, which indicates a relatively small $^{56}$Ni mass (see
\S\ref{s:model} for more information on the mass of the ejecta).
The line profiles of 
\ion{Mg}{1}] $\lambda$4571, [\ion{O}{1}] $\lambda\lambda$6300, 6364,
and [\ion{Ca}{2}] $\lambda\lambda$7291, 7324 in the spectra of \ig\ in
Figure~\ref{f:late_spec} are quite strong and complex;
Figure~\ref{f:OMG_comparison}
illustrates them in detail.  
The \ion{Mg}{1}] and [\ion{O}{1}] lines at all epochs appear
to have a broad peak, punctuated by multiple local maxima. \ion{Mg}{1}] has peaks
near $-$2000~\kms\ and at $-$1000~\kms\ relative to the systemic velocity of
\object{NGC 7424} \citep[939~\kms;][]{Koribalski04}.  This line may also have another bump near $-$3000~\kms\ in the
blue wing of the main feature. The [\ion{O}{1}] emission is even more complex, 
with local
peaks near $-$3000~\kms, $-$1700~\kms, 150~\kms, and 1200~\kms\ (again relative to the systemic
velocity of \object{NGC 7424}).  
The [\ion{Ca}{2}]
$\lambda\lambda$7291, 7323 emission in Figure~\ref{f:OMG_comparison} has a broad peak
which is blueshifted by \about1000~\kms\ in all three epochs. 

Figure~\ref{f:OMG_comparison} also shows the temporal evolution of the \ion{Mg}{1}]
$\lambda$4571 and [\ion{O}{1}] $\lambda\lambda$6300, 6364 line
profiles. One of the most notable changes is the
decrease in relative flux in the far wings of the \ion{Mg}{1}] line near
$\pm$5000~\kms. As mentioned above, the feature near $-$3000~\kms\ appears to decrease 
in strength with time. This decrease of
relative flux with time near $-$3000~\kms\ is also seen in the
[\ion{O}{1}]
feature.  Finally, we note that there is also an
increase in the flux in the wings of the [\ion{Ca}{2}] feature near
$\pm$2000--5000~\kms.

The complex structure of the \ion{Mg}{1}]
$\lambda$4571 and [\ion{O}{1}] $\lambda\lambda$6300, 6364 lines has
been seen previously in another well-studied Type IIb, SN~1993J
\citep{Matheson00}. They found that the [\ion{O}{1}] and \ion{Mg}{1}]
line profiles resulted from clumps in the ejecta
while the [\ion{Ca}{2}] lines did not.
They also pointed out that this is consistent with the
explosion scenario first proposed for SN~1987A by \citet{Li92,Li93}, in
which the emission from oxygen comes from clumps
of newly synthesized material while the calcium emission originates
mainly in preexisting uniformly distributed material.
\citet{Filippenko89} found similar irregularities in the [\ion{O}{1}]
$\lambda\lambda$6300, 6364 lines of the SN~Ib~1985F.  They suggested that
the spectral peculiarities come either from Rayleigh-Taylor fingers of
higher-speed material or from local density enhancements.

The [\ion{O}{1}] doublet of \ig\ somewhat resembles
that of the peculiar SN~Ib~2005bf \citep{Modjaz08} in that \ig\
may also show a double-peaked profile with one of the peaks
near zero velocity and the other blueshifted by a few thousand~\kms.
However, the \ion{Mg}{1}] line in \ig\ at all epochs has a broad
profile with a complex top and
does not seem to exhibit the same double-peaked nature of the
[\ion{O}{1}] emission. This differs from the results of
\citet{Foley03} for the energetic broad-lined SN~Ic~2002ap, which
showed identical [\ion{O}{1}] and \ion{Mg}{1}] profiles. In addition,
the flux in the \ion{Mg}{1}] line appears to be more
centrally concentrated in \ig\ compared to SN~2002ap.

It has been proposed that the double-peaked [\ion{O}{1}]
profile of SN~2005bf was caused by a unipolar blob or jet \citep{Maeda07} while
its polarization signature was attributed to a possible ``tilted jet''
passing through an asymmetric photosphere \citep{Maund07b,Maund07}. Both our models
of late-time spectra of \ig\ and the similarity of the profiles in
our late-time spectra to those of SN~1993J \citep{Matheson00} indicate
clumpy, oxygen-rich
ejecta as well (see \S\ref{s:model} for more information regarding the
models and ejecta). Thus,
\ig\ probably has large-scale asymmetry in its ejected oxygen, either in
the form of a blob or jet (tilted or otherwise), while the
distributions of magnesium and calcium appear more uniform (though the
magnesium seems only marginally more uniform than the oxygen). This
is supported by \citet{Fransson89}, who predicted that magnesium and
calcium should be found closer to the core of the explosion than oxygen.

\citet{Filippenko02} stated
that the \ion{Mg}{1}] $\lambda$4571 feature
was the strongest (relative to [\ion{O}{1}] $\lambda\lambda$6300, 6364) ever seen
in a SN. We investigate this
claim more closely and the results are shown in Table~\ref{t:lines},
where we list the relative line strengths for each nebular spectrum
of \ig\ along with some of the values from \citet[][Table 6]{Foley03}
and values calculated from other SN spectra in the literature.
To compute the relative line strengths for \ig\ we subtracted a local
continuum from each spectral feature and then integrated the flux in
each line.  This is the same procedure presented by
\citet{Foley03}.


\begin{deluxetable*}{lcccccc}
\tablecaption{Integrated Fluxes Relative to [\ion{Ca}{2}]
  $\lambda\lambda$7291, 7324\label{t:lines}}
\tablewidth{0pt}
\tablehead{
\colhead{Supernova} &
\colhead{Day\tablenotemark{a}} &
\colhead{\ion{Mg}{1}]} &
\colhead{[\ion{O}{1}]} &
\colhead{\ion{Ca}{2} near-IR} &
\colhead{Mg/O Ratio} \\
 & &
\colhead{4571 \AA} &
\colhead{6300, 6364 \AA} &
\colhead{8498, 8542, 8662 \AA} & }
\startdata
\ig (IIb)\tablenotemark{b}         & 256 & 0.38 & 1.77 &\nodata\tablenotemark{c}& 0.21 \\
\ig (IIb)                          & 309 & 0.74 & 2.06 & 0.30 & 0.36 \\
\ig (IIb)                          & 340 & 0.83 & 1.99 & 0.57 & 0.42 \\
SN~2008D (Ib)\tablenotemark{d}     & 109 & 0.34 & 1.02 & 5.61 & 0.33 \\
SN~2006aj (Ic-BL)\tablenotemark{e}  & 153 & 1.19 & 4.39 & 1.37 & 0.27 \\
SN~2003jd (Ic-BL)\tablenotemark{f}  & 367 & 0.73 & 1.58 & 0.80 & 0.46 \\
SN~1993J (IIb)\tablenotemark{g}    & 298 & 0.27 & 1.79 & 0.36 & 0.15 \\
SN~2002ap (Ic)\tablenotemark{h}    & 242 & 0.72 & 2.17 & 0.91 & 0.33 \\
SN~2002ap (Ic)\tablenotemark{h}    & 274 & 0.63 & 2.01 & 0.66 & 0.31 \\
SN~2002ap (Ic)\tablenotemark{h}    & 336 & 1.00 & 2.29 & 0.45 & 0.43 \\
SN~2002ap (Ic)\tablenotemark{h}    & 386 & 0.87 & 2.19 & 0.46 & 0.40 \\
SN~1998bw (Ic)\tablenotemark{h}    & 215 & 0.23 & 1.47 & 0.41 & 0.15 \\
SN~1994I (Ic)\tablenotemark{h}     & 146 & 0.26 & 0.97 & 0.84 & 0.26 \\
SN~1985F (Ib/c)\tablenotemark{h}   & \about240  & 0.26 & 3.15 & 0.56 & 0.08 \\
\enddata
\tablenotetext{a}{Days since explosion.}
\tablenotetext{b}{Spectrum previously published by \citet{Maund07}.}
\tablenotetext{c}{Spectrum ends at 8600 \AA.}
\tablenotetext{d}{Spectrum previously published by \citet{Modjaz09}.}
\tablenotetext{e}{Spectrum previously published by \citet{Mazzali07:06aj}.}
\tablenotetext{f}{Spectrum previously published by \citet{Valenti08}.}
\tablenotetext{g}{Spectrum previously published by \citet{Filippenko94}.}
\tablenotetext{h}{Values from \citet[][Table 6]{Foley03}.}
\end{deluxetable*}

Although it has been pointed out that this method of line-strength measurement is somewhat
inaccurate \citep{Foley03}, it does seem to indicate that the
\ion{Mg}{1}] line of \ig\ grows with time with respect to the
[\ion{Ca}{2}] $\lambda\lambda$7291, 7324 doublet and the [\ion{O}{1}] doublet.
Relative to the [\ion{Ca}{2}] feature, the \ion{Mg}{1}] emission from
\ig\ is somewhat weaker than that of SN~2002ap at similar epochs.
The other SNe that also appeared in \citet[][Table 6]{Foley03} all
have about the same  \ion{Mg}{1}] strength, which is lower that that
of either SNe~2001ig or 2002ap.  However, we also perform this
measurement on a few SNe from the literature (see Table~\ref{t:lines}) and find that 
while SNe~2008D \citep{Modjaz09} and 1993J \citep{Matheson00} have 
comparable \ion{Mg}{1}] strength, this feature is quite strong in 
the broad-lined Type~Ic (Ic-BL) SNe~2006aj \citep{Mazzali07:06aj}
and 2003jd \citep{Valenti08}.

Again comparing \ig\ to the SNe that appeared in \citet[][Table
6]{Foley03}, it has a Mg/O ratio comparable to that of SNe~1994I and 2002ap,
but larger than that of SNe~1998bw and 1985F.  In addition, the Mg/O
ratio of \ig\ is similar to that of SNe~2008D, 2006aj, and 2003jd
(which in fact has quite a large ratio), but significantly greater than
that of SN~1993J.  This confirms the claim of \citet{Filippenko02}
that \ig\ had the strongest nebular O/Mg emission-line ratio
of any SN observed at that time.  Table~\ref{t:lines} also
seems to indicate that it is not uncommon for 
core-collapse SNe to have relatively large Mg/O ratios, yet what
distinguishes between the SNe that do or do not show this is unclear.


%


\section{Spectral Models}\label{s:model}

By several months after the explosion of a SN, the densities in the
ejecta are sufficiently low from expansion that the opacity drops below unity and the gas obeys
nebular physics. At this time it is heated by radioactive decay of 
$^{56}$Ni and cooled by line emission.
The resulting line profiles map rather directly the distribution of the heating
material and can be used to derive several properties including
$^{56}$Ni mass, mass of the emitting elements, and morphology of the
explosion \citep{Mazzali05}.

In order to derive accurate models, we need our spectra to be as
spectrophotometrically well calibrated as possible.  While the
relative fluxes of our data are accurate because of our
flux-calibration routines (see \S\ref{s:obs}), the absolute fluxes may
be off due to imperfect observing conditions.  To address this, we
obtained publicly available photometric data on \ig\ from ESO and
the Space Telescope -- European Coordinating Facility (ST-ECF) Science
Archive Facility\footnote{http://archive.eso.org/}.  The observations
were single 60~s Bessell \citep{Bessell90} $R$-band images taken with the FORS1
spectrograph mounted on the ESO 8.2-m Very Large Telescope Melipal and 
Antu \citep{Appenzeller98}.

We calibrated both images, observed on 2002~Oct.~12 and 2002~Dec.~8,
to the United States Naval Observatory B1.0 (USNOB) Catalog with
three comparison stars and measured magnitudes for \ig\ of
17.99$\pm$0.08 and 19.13$\pm$0.14, respectively.  In addition
to our stated statistical uncertainty, we adopt a systematic uncertainty
of \about0.25~mag based on comparisons of USNOB magnitudes to Bessell
$R$-band magnitudes obtained as part of the Lick Observatory Supernova 
Search \citep[LOSS;][]{Filippenko01} 
photometry follow-up program. We then synthesized Bessell 
$R$-band  magnitudes from our late-time
spectra (taken on 2002~Oct.~8 and 2002~Nov.~8) and compared them to
the magnitudes derived from the photometry, assuming a linear decline
in magnitude during the last three months of 2002.  In order for our
spectra to match the photometry, and thus for the absolute flux to be
accurate, our spectra from 2002~Oct.~8 and 2002~Nov.~8 must be scaled
up by factors of 4.41 and 1.29, respectively.  Due to the photometric measurements alone, 
the uncertainty in the absolute flux of each of these two spectra is \about25\%.

We modeled both late-time spectra of \ig\ using a code that computes 
nebular-line
emission from a gaseous cloud. The gas is heated by the deposition of 
$\gamma$-rays and positrons produced by the decay of radioactive 
$^{56}$Ni (which is produced in the SN explosion itself) into
$^{56}$Co and then into
stable $^{56}$Fe. The energy thus deposited in the gas is thermalized
collisionally, which leads to excitation and ionization of the gas. Heating is
balanced by cooling, which takes place via line emission. Most of the emission
lines are forbidden, but some permitted transitions (e.g., the
\ion{Ca}{2} near-IR triplet) are
also active. The code is based on the original description by \citet{Axelrod80}.
The propagation of the $\gamma$-rays is followed with a Monte Carlo 
procedure \citep{Cappellaro97}.
Level populations are computed assuming non-local thermodynamic equilibrium
(non-LTE). A more detailed description is provided by \citet{Mazzali07:02ap}.

Modeling the spectra also requires knowledge of the reddening of, and
distance to, the SN. As discussed in
\S\ref{ss:early}, the reddening within \object{NGC 7424} is negligible and
we have only an upper limit from our data, so we adopt a total
reddening of $E\left(B-V\right)=0.011$~mag from \citet{Schlegel98} for the
models. For the distance to \ig\
we have used a recession velocity of 939~\kms\
\citep[$z=0.003132$;][]{Koribalski04}
throughout this paper, yielding a distance modulus
of $m-M=30.5$~mag with
$H_0 = 73$~\kms~\perMpc\ \citep{Riess05}, assuming the entire
recession velocity is due to the Hubble flow.  However, there is some
uncertainty in this value due to the relatively small distance to
\object{NGC 7424}.  A search of the literature reveals a range in
distance moduli for this galaxy of 30.2--30.7~mag using various
independent techniques
\citep[HyperLEDA\footnote{http://leda.univ-lyon1.fr/};][private communication\footnote{Using
 the 2MASS Redshift Survey (2MRS) galaxy catalog to model bulk flows
 in the local Universe.}]{Tully88,Mould00,Meurer06,Davis09}.
The value of 30.5~mag that we use here is a reasonable average
as long as we keep in mind that it has an uncertainty
of \about0.3~mag which leads to an uncertainty of \about30\% in the absolute flux of each 
late-time spectrum.  When this is combined in quadrature with the uncertainty from the
photometric measurements, the total uncertainty in the absolute flux of
each of our late-time spectra is \about39\%.

%
%
%
%
%

\begin{figure*}
\epsscale{1.2}
\rotatebox{90}{
\plottwo{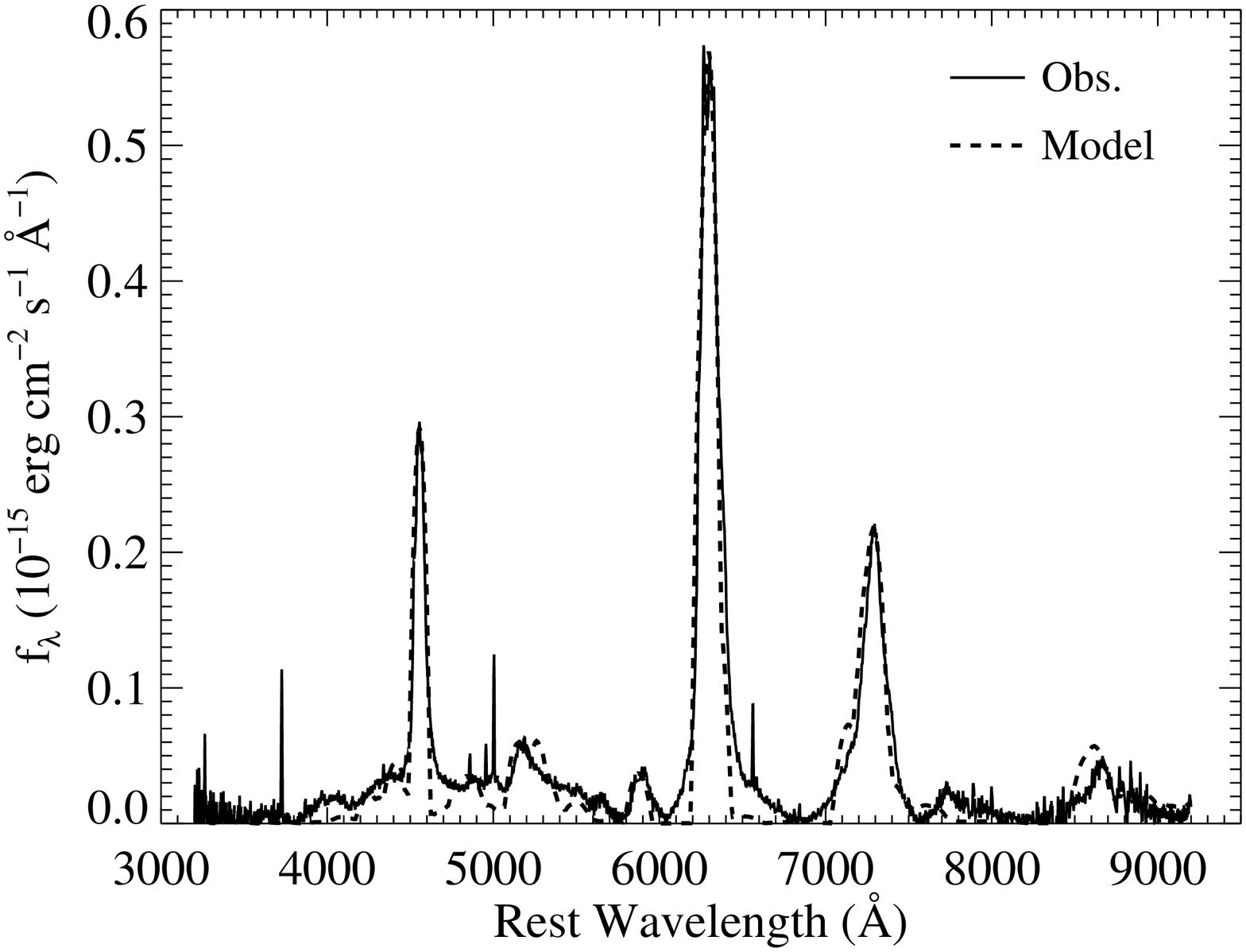}{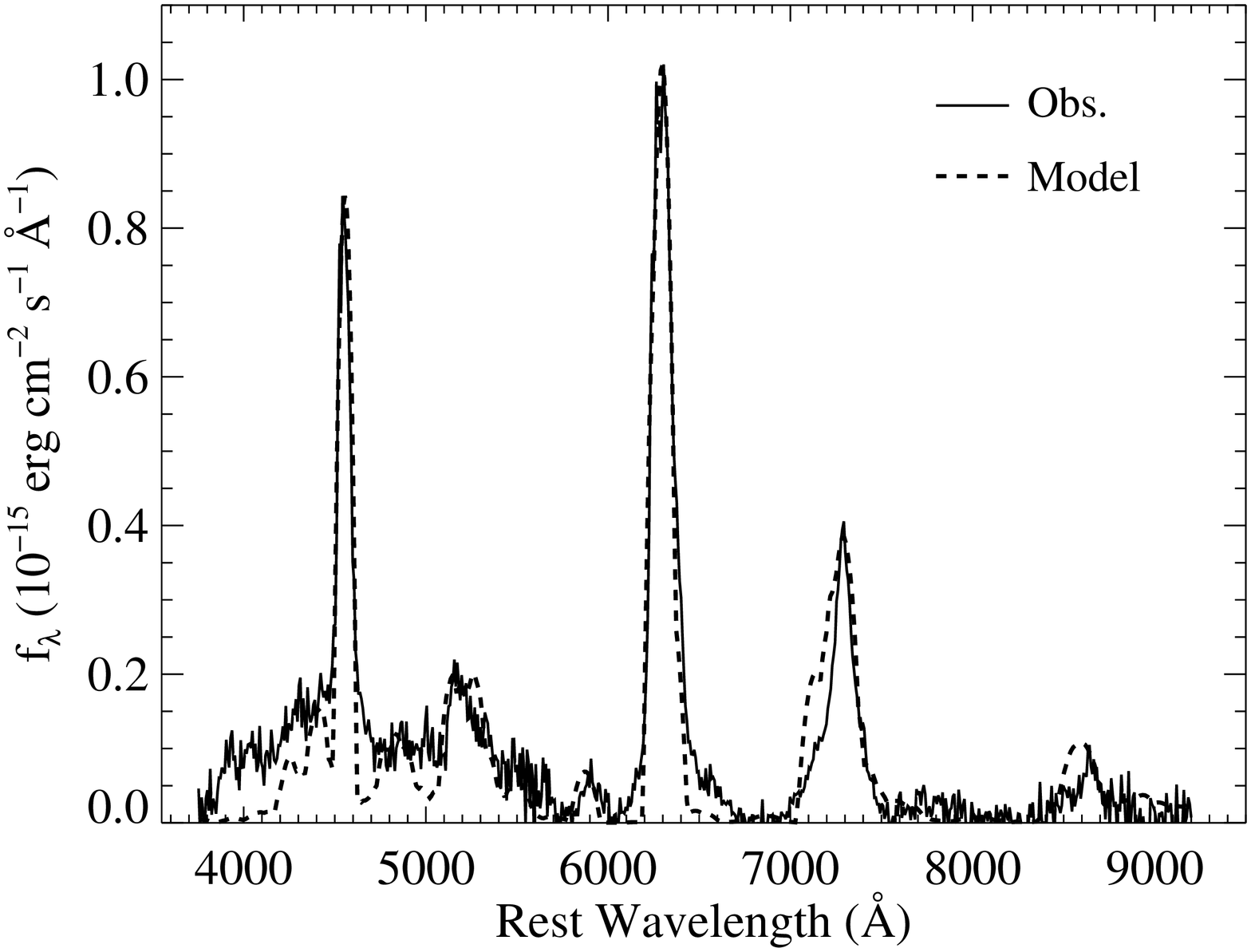}}
\caption{The model and observed spectra of \ig\ 
from 2002~Oct.~8 (309 days after explosion, top) and 2002~Nov.~8 (340 days 
after explosion, bottom). The recession velocity of
the host galaxy NGC 7427 has been removed from the spectra. The model
accurately reproduces the major spectral features seen in the data.}\label{f:model}
\end{figure*}

The late-time spectra and models are shown in Figure~\ref{f:model};
both models appear to fairly accurately
reproduce the major features observed in the data.  However, the
emission near 4000~\AA\
and the \ion{O}{1} feature near 7700~\AA\ seem somewhat suppressed in the models. 
In addition, the intensity of the \ion{Ca}{2} feature near 8600~\AA\ in the models
is somewhat larger than that of the observed spectrum, leading to a small
inconsistency in the position of the combined \ion{Ca}{2} plus
[\ion{C}{1}] $\lambda$8727 emission.

Here we use the one-zone version of the model described by
\citet{Mazzali07:02ap}. It requires an emitting nebula with an outer
velocity of 4300~\kms, as
determined from reproducing the width of the strongest emission lines
([\ion{O}{1}], \ion{Mg}{1}], and [\ion{Ca}{2}]). The spectrum is powered by 
\about0.13~\msun\ of $^{56}$Ni. This is determined from the combination of 
the intensity of the [\ion{Fe}{2}] emission feature near 5200~\AA\ and the 
requirement that the energy emitted in the radioactive decay chain of $^{56}$Ni 
excites other ions and reproduces the observed emission in other lines. 
This is a slightly larger value than what was derived for SN~1993J
\citep[see][and references therein]{Swartz93}, but it is
still fairly typical for core-collapse SNe.

The total mass enclosed within the outer velocity of 4300~\kms\ is
\about1.15~\msun.  Interestingly, our models of \ig\ are quite similar to that of the
SN~Ic~1994I which underwent significant mass loss, most likely in a binary
interaction, and which also ejected \about1~\msun\ of material \citep{Sauer06}.
\citet{Nomoto94} showed the progenitor star of SN~1994I to be \about15~\msun\ initially,
which suggests that the progenitor of \ig\ was also initially \about15~\msun.

Furthermore, \citet{VanDyk02} and \citet{Maund04} found that both the
companion and the progenitor of SN~1993J had intial masses of
\about15~\msun\ and that the companion of
SN~1993J is currently 22~\msun.  \citet{Ryder06} discovered that the
companion of \ig\ is currently 10--18~\msun, so it seems
likely that the progenitors of \ig\ and SN~1993J (and possibly even
SN~1994I) had similar initial
masses, and that their companions have similar masses as well.
The differences in the
companions' masses (as well as physical separations from the
progenitors) probably led to slightly different mass-loss and interaction
histories which gave rise to the spectroscopic differences observed.

Table~\ref{t:model} shows the complete mass composition of the models. 
The upper and lower error bars represent the change in the
derived abundance after increasing and decreasing the measured flux by 39\%
to reflect the flux-calibration uncertainties mentioned above.
As in the nebular spectra of all other stripped-envelope SNe, and unlike 
the case in
SNe Ia, [\ion{Fe}{3}] lines are missing (in particular a strong feature near
4800~\AA). This is best reproduced by assuming fairly significant clumping in 
the ejecta, which is a signature of an aspherical explosion
\citep[e.g.,][]{Mazzali07:02ap}.  \citet{Maund07} use
spectropolarimetric observations of \ig\ to come to the same
conclusion regarding the asphericity.

The dominant element in the ejecta by mass is oxygen (0.81~\msun). The derived carbon abundance is quite small 
(0.04~\msun); otherwise, the emission near 8500~\AA\ would become too strong. 
We point out that a small C/O ratio is not unusual in stripped-envelope SNe.
In addition, hydrogen may be present only marginally at these low velocities, as deduced from
the almost complete absence of any emission (as seen in
Figure~\ref{f:late_spec}).

As mentioned in \S\ref{ss:late}, the late-time spectra of \ig\ have
an unusually large Mg/O intensity ratio, comparable to that of
SN~2002ap. However, the Mg/O ratio by mass of \ig\ is \about0.014 (see
Table~\ref{t:model}) while that of SN~2002ap is \about0.032 \citep{Mazzali07:02ap}.  
This difference in masses is consistent with the fact that the envelope of SN~2002ap was
extremely stripped before explosion,
since it showed no spectroscopic evidence of hydrogen or helium
\citep{Foley03}.  On the other hand, the envelope of \ig\ was only partially
stripped prior to core collapse since we observed a significant
amount of hydrogen and helium at early times.


\begin{deluxetable}{lcc}
\tablecaption{Mass Composition of the Models\tablenotemark{a}\label{t:model}}
\tablewidth{0pt}
\tablehead{
\colhead{Element} &
\colhead{Mass (\msun)} &
\colhead{Mass (\msun)}  \\
 & 
\colhead{Day 309} &
\colhead{Day 340} }
\startdata
C    &    $2.0^{+0.0}_{-0.8}\times10^{-2}$  &  $5.0^{+0.0}_{-1.0}\times10^{-2}$  \\
O    &    $8.1^{+2.0}_{-2.1}\times10^{-1}$  &  $8.2^{+1.8}_{-1.7}\times10^{-1}$  \\
Na   &   $7.0^{+0.0}_{-1.5}\times10^{-5}$  &  $1.0^{+0.2}_{-0.0}\times10^{-4}$  \\
Mg   &   $1.1^{+0.2}_{-0.0}\times10^{-2}$  & $1.1^{+0.2}_{-0.0}\times10^{-2}$   \\
Si   &    $1.0^{+0.0}_{-0.0}\times10^{-1}$   &  $1.0^{+0.0}_{-0.0}\times10^{-1}$  \\
S    &    $3.0^{+0.0}_{-0.0}\times10^{-2}$   &  $3.0^{+0.0}_{-0.0}\times10^{-2}$  \\
Ca   &   $3.1^{+0.8}_{-0.9}\times10^{-2}$  &  $4.0^{+1.2}_{-0.9}\times10^{-2}$   \\
$^{56}$Ni   &    $1.5^{+0.3}_{-0.3}\times10^{-1}$   &  $1.1^{+0.2}_{-0.3}\times10^{-1}$  \\
\hline
Total         &    $1.2^{+0.3}_{-0.3}$   & $1.2^{+0.2}_{-0.2}$ \\
\enddata
\tablenotetext{a}{The upper and lower error bars represent the change in the
derived abundance after increasing and decreasing the measured flux by 39\%
to reflect the flux-calibration uncertainties mentioned in \S\ref{s:model}.}
\end{deluxetable}


\section{Conclusions}\label{s:conclusions}
In this paper we have presented and analyzed
optical spectra of \ig. One week after explosion, the SN defied simple
spectral classification, but by two weeks after explosion \ig\
appeared to be part of the Type~IIb subclass of SNe.
This was confirmed as the SN went
through a transition from Type~II to Type~Ib during the first few
months of observations.

By nine months after explosion, \ig\ had entered the nebular phase, 
revealing some of the strongest \ion{Mg}{1}] $\lambda$4571 and
[\ion{O}{1}] $\lambda\lambda$6300, 6364 features ever observed in a SN;
its Mg/O intensity ratio is one of the largest ever seen as well.

We derive models of \ig\ from our spectra taken 309 and 340 days
after explosion, showing that the majority of the inner ejecta 
(below 4300~\kms) were
in the form of oxygen and a significant fraction (nearly 0.13~\msun) in
the form of $^{56}$Ni.  Additionally, there appears to be a distinct lack
of hydrogen in the inner ejecta.  We also find that the total mass of
this inner ejecta was \about1.15~\msun, suggesting
that the progenitor of \ig\ was a relatively low-mass star
(\about15~\msun).


\begin{acknowledgments}
We are grateful to the staffs at the Keck Observatory, ESO, and Las
Campanas Observatory for their support.
We thank the following for their assistance with some of the
observations and data reduction: F. Barrientos, R. Carlberg, M. Gladders,
S.~W. Jha, T. Matheson, J.~L. Prieto, and B.~Leibundgut. We especially thank J. Maund
and his collaborators for allowing us to reproduce one of their
spectra.
Some of the data presented herein were obtained at the W. M. Keck
Observatory, which is operated as a scientific partnership
among the California Institute of Technology, the University of California,
and the National Aeronautics and Space Administration;
the observatory was made possible by the generous financial
support of the W. M. Keck Foundation. The authors wish to recognize
and acknowledge the very significant cultural role and reverence
that the summit of Mauna Kea has always had within the indigenous
Hawaiian community; we are most fortunate to have the opportunity
to conduct observations from this mountain.  ESO VLT data were acquired
under programs 68.D-0571(A), 69.D-0438(A), and 170.A-0519(A), and La
Silla data under program 68.A-0443.
A.V.F.'s group is supported by the National Science Foundation 
through grant AST--0607485.
A.C. is supported by grants P06-045-F ICM de MIDEPLAN, Basal CATA PFB
06/09, and FONDAP No. 15010003.

\end{acknowledgments}

\bibliographystyle{fapj}
\bibliography{astro_refs.bib}

\end{document}